      [1999/12/01 v1.4c Il Nuovo Cimento]
\documentclass{cimento}

\usepackage{graphics,graphicx}
\usepackage{epsfig}

\title{GRIPS and its strong connections to the GeV and TeV bands}
\author{O.~Tibolla\from{ins:x},
K.~Mannheim\from{ins:x},
A.~Paravac\from{ins:x},
J.~Greiner\from{ins:y}\\
        \atque
G. Kanbach\from{ins:y}}
\instlist{\inst{ins:x} Institut f\"ur Theoretische Physik und Astrophysik, Universit\"at W\"urzburg, Germany; Omar.Tibolla@astro.uni-wuerzburg.de

  \inst{ins:y} Max Plank Institut f\"ur Extraterrestrische Physik, Garching, Germany
}
\PACSes{\PACSit{07.85.-m} {07.87.+v} {98.70.Sa} {98.58.Mj}}

\begin{document}

\maketitle

\begin{abstract}

{\it GRIPS} is planned to be the next great satellite-born survey mission lead by Europe; it will look into the cosmos with unprecedented accuracy in several bands of the EM spectrum (infrared, X-rays, MeV gamma-rays); in particular in gamma-rays, {\it GRIPS} will be able to bridge the so-called MeV gap and to answer several questions brought forth by GeV-TeV gamma-ray observations. We will discuss here several connections to GeV-TeV gamma-ray astrophysics, focussing in particular to show how {\it GRIPS} will be crucial in revealing the origin of cosmic rays.

\end{abstract}

\section{{\it GRIPS}}
The Gamma-Ray Imaging, Polarization and Spectroscopy ({\it GRIPS}) mission \cite{web} consists of a Gamma-ray monitor (GRM), an X-ray monitor (XRM) and an Infrared telescope (IRT):

- the IRT is an IR telescope with a 1 meter mirror and it will be a copy of {\it Euclid} \cite{euclid}; in addition, in IRT focal point there is set a GROND \cite{grond} like system with 7 detectors, with a $10' \times 10'$ field of view;

- the XRM is re-using {\it eROSITA} \cite{erosita} technology; it will work in the energy range 0.1 - 10 keV with a spatial resolution of $\sim 30''$;

- the GRM will work in the energy range 200 keV - 80 MeV with a spatial resolution of $\sim 1^{\circ}$ and it will also provide polarization ($\sim 1$\%).

\begin{figure}
\begin{center}
\hbox{
\psfig{figure=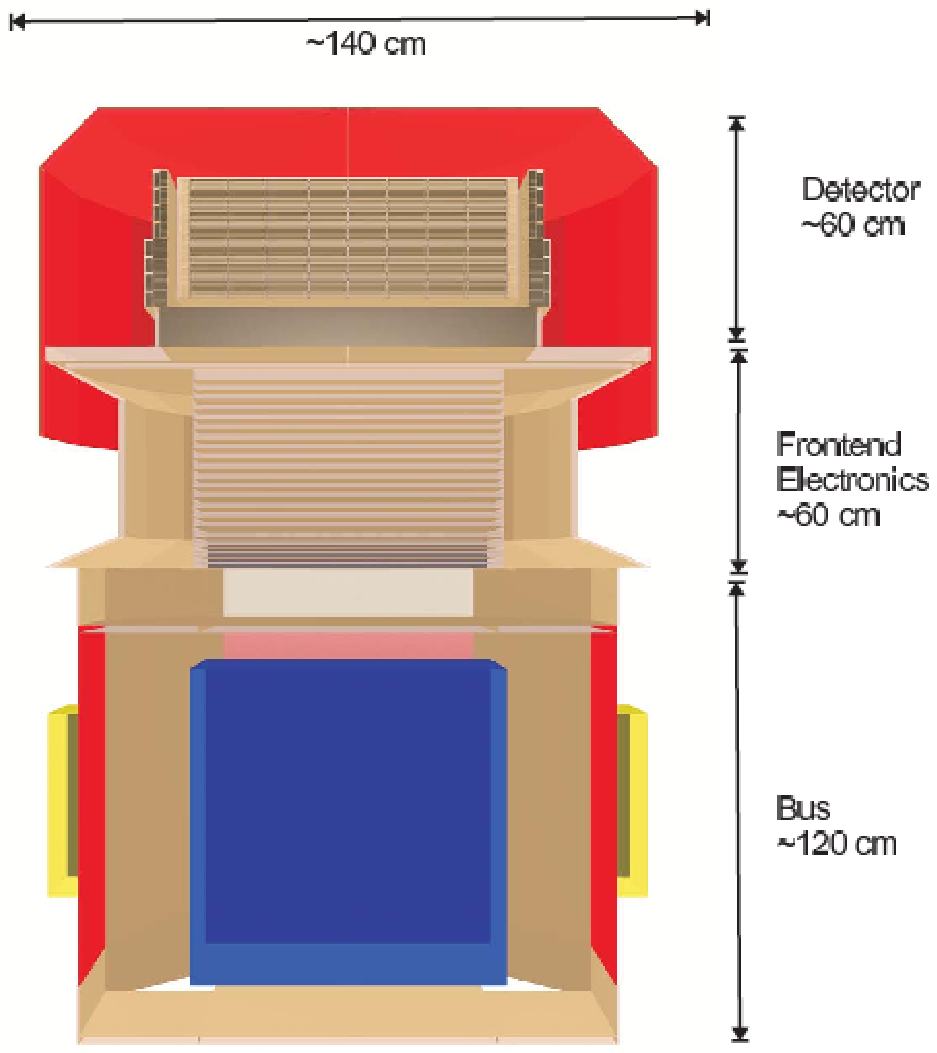,height=6.0cm,angle=0}
\hspace{2.5cm}
\psfig{figure=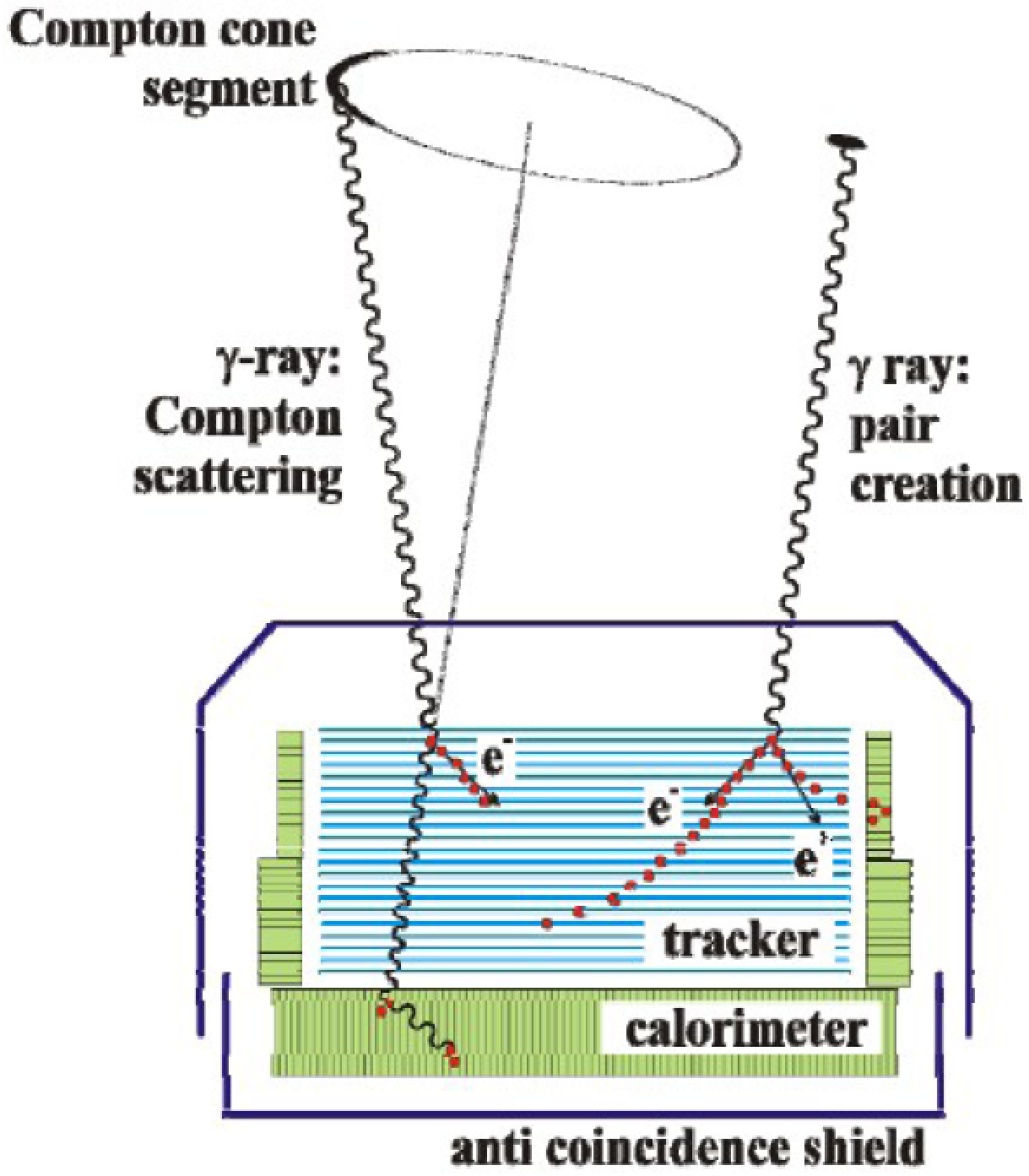,height=6.0cm, angle=0}
\vspace{-1cm}
}
\end{center}
\caption{\footnotesize
{{\it Left panel}: Schematic representation of the GRM bus.
{\it Right panel}: Detail on GRM working principle and its schematic representation.
}
}
\label{fig1}
\end{figure}


These 3 instruments will be set in orbit onboard 2 different satellite buses, however, respecting the {\it Soyuz} weight limits (the total weight of GRIPS is 5 tons, while the Soyuz-Fregat payload limit is 5.2 tons), it will require one single {\it Soyuz} launcher. Fig. \ref{fig1} shows a schematic representation of the bus carrying GRM (while IRT and XRT will fly onboard the second satellite bus), zooming into GRM working principle.

GRM is a combined Compton and pair conversion telescope, consisting of 3
different parts:

- A Tracker (TKR), made of Silicon double-side detectors (DSD), where the initial interaction takes place and the secondary charged particles are tracked; the TKR consists of 4 towers, each tower consists of 64 layers spaced 5 mm from each other (also a configuration with 3 mm spacing has been under investigation) and each layer consists of $4 \times 4$ DSD wafers of
$10 \times 10$ cm$^2$ each.

- A Calorimeter (CAL), made of LaBr$_3$, where the energy of the secondary particles is converted into light signal.

- An Anti-coincidence Shield (ACS), made of plastic scintillator (Ne110), that will shield the instrument from charged particles; note that the efficient background rejection, provided by the ACS, is a crucial point for a MeV telescope.

A detailed description of GRIPS mission can be found in \cite{greiner} and in the original project \cite{greiner2}. 

\section{Connections to higher energies}

Looking from a GeV and TeV gamma-rays perspective, we list several scientific topics  that show remarkable connections, putting emphasis on the {\it origin of the Cosmic Rays}.

\subsection{The MeV-gap}

With the advent of {\it Fermi LAT} thousands of sources have been detected in the GeV gamma-ray band ({\it e.g.} \cite{1fgl}) and at keV energies thousands of sources were already known: instead in the MeV band only $\sim$ 40 sources are known \cite{comptel}.

\begin{figure}
\begin{center}

\psfig{figure=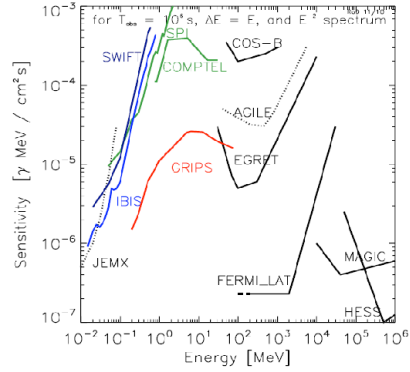,height=5.0cm,angle=0}
\vspace{-0.5cm}

\end{center}
\caption{\footnotesize
{GRM-Grips sensitivity for $10^6$s observation time, approximating its energy dispersion as $\Delta E = E$.
}
}
\label{fig2}
\end{figure}

Fig. \ref{fig2} shows GRM-GRIPS sensitivity for $10^5$s observation time; it shows that at 1 MeV GRM-GRIPS is 40 times more sensitive than {\it Comptel/INTEGRAL} and this implies the detection of more than 1000 sources in a 1 year survey! Fig. \ref{fig2} underlines how the MeV band is a poorly studied region of the electromagnetic spectrum, how GRM-GRIPS is complementary with gamma-ray and with soft X-ray experiments and how GRM-GRIPS will be able to bridge the gap between these two bands.

\subsection{Gamma-ray Bursts}

With the advent of {\it Swift}, {\it Fermi} GBM and {\it Fermi} LAT, gamma-ray bursts (GRBs) science is living a golden age.
{\it GRIPS} will operate in survey-mode and, thanks to its great field of view ($\sim 160^{\circ}$), will be a great GRB monitor (with more than 600 GRBs expected per year).

\begin{figure}
\begin{center}
\hbox{
\psfig{figure=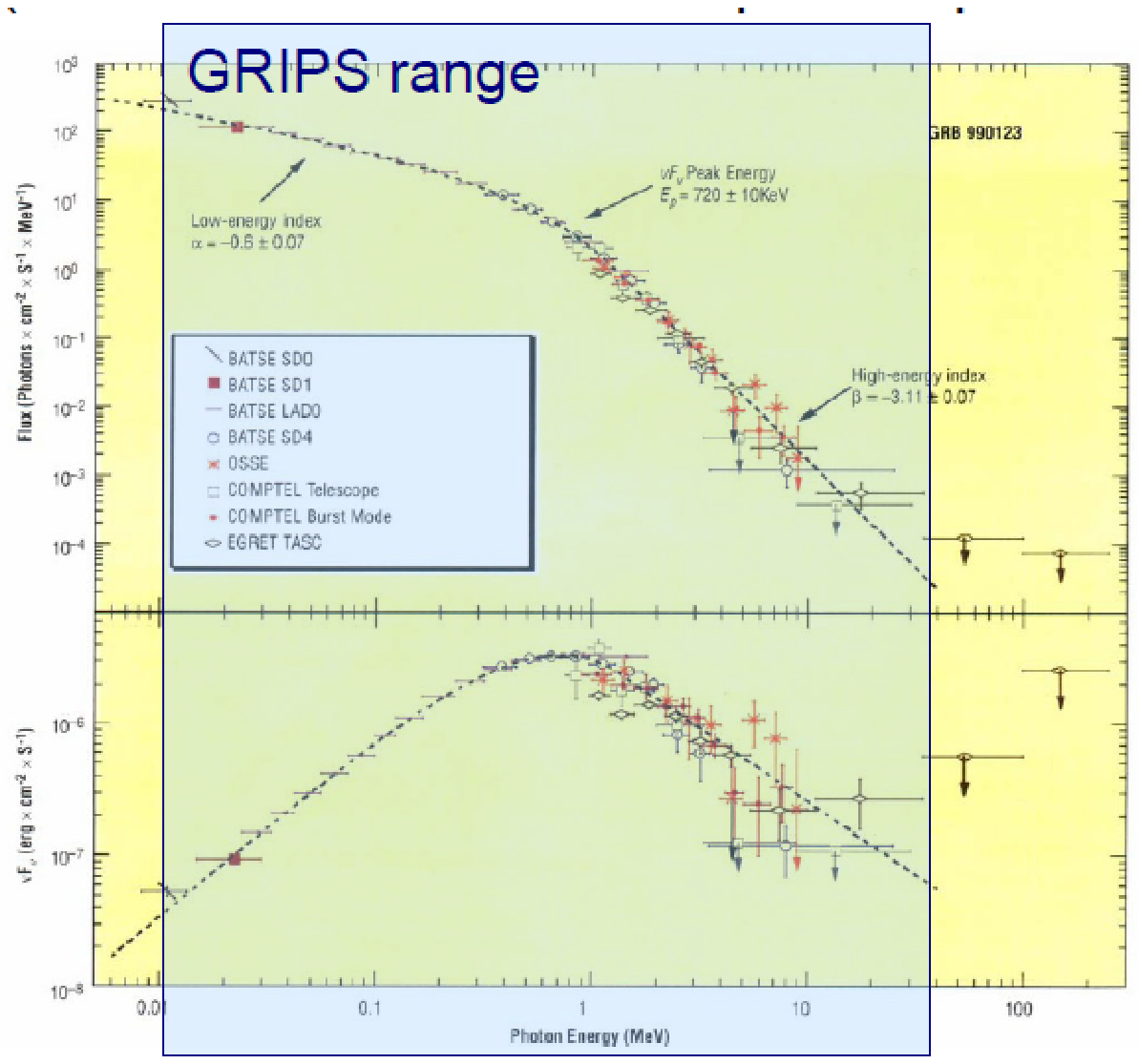,height=5.0cm,angle=0}
\hspace{0.5cm}
\psfig{figure=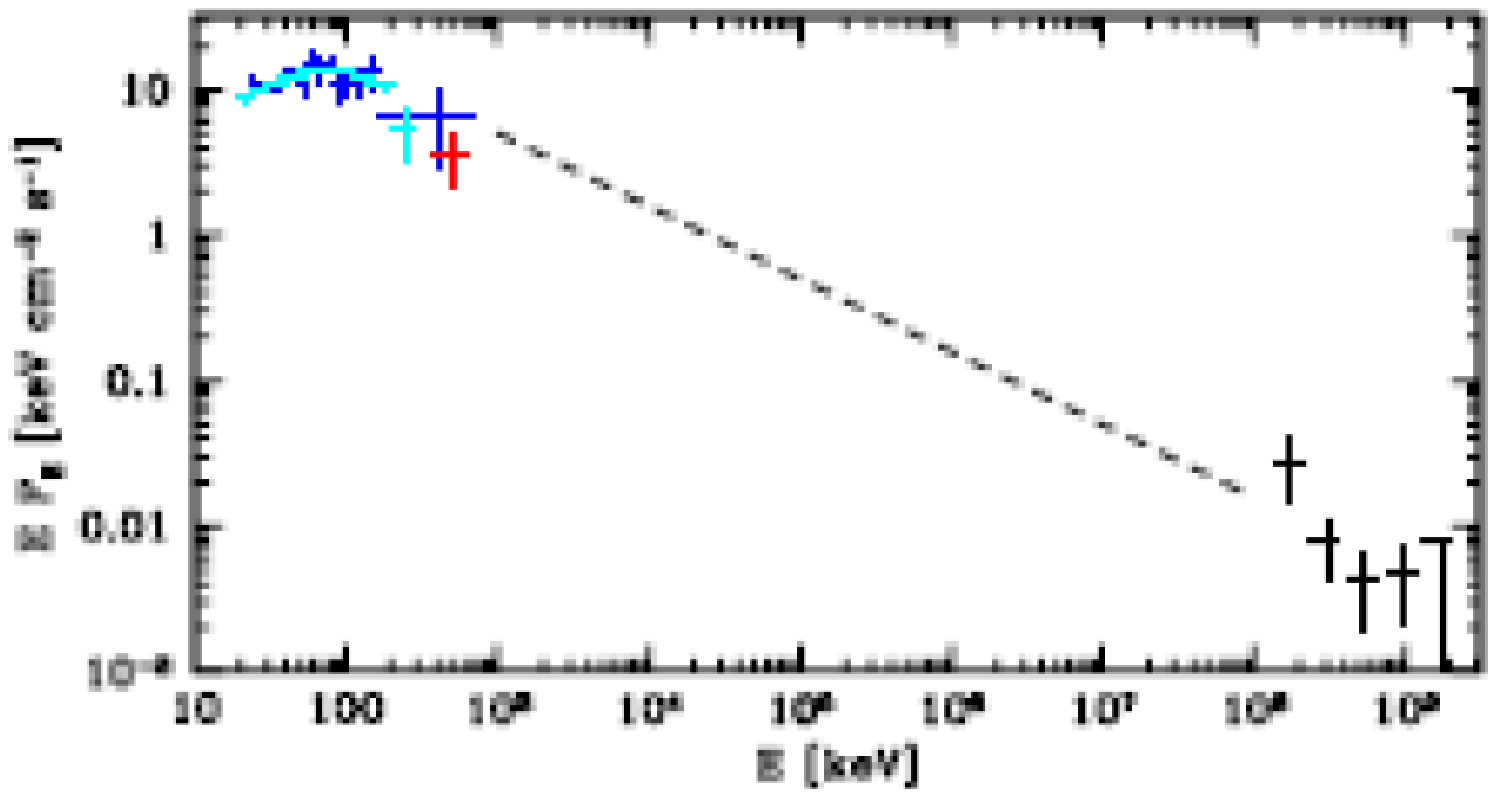,height=4.0cm,angle=0}
\vspace{-1cm}
}
\end{center}
\caption{\footnotesize
{{\it Left panel}: The example of GRB 990123 \cite{990123} with GRM-Grips sensitivity overlapped.
{\it Right panel}: The example of Cygnus X1 taken from \cite{x1}.
}
}
\label{fig3}
\end{figure}

Moreover the MeV band is extremely important for GRB science, since the $\nu$F$\nu$ energy peak is at MeV energies ({\it e.g.} \cite{band}), as shown in fig. \ref{fig3}.

Finally, in case we would find density columns of $10^{25}$ cm$^{-2}$ and higher (structure formations at $z=20$) we could study resonance photon absorptions, such as the Pigmy Dipole Resonance (PDR; produced by the resonance capture of photons on nuclear levels with energy of 3-9 MeV) and the Giant Dipole Resonance (GDR; process that is naively described as oscillation of the neutron fluid relative to protons, at 15-25 MeV), while the Delta-resonance (at $\sim$ 325 MeV) is out of range; see \cite{GRB930131} for more details.

\subsection{Supra-thermal/Non-thermal emission}

The thermal components finish at maximum at some hundreds of keV; and the transition from thermal to non-thermal is crucial for many astrophysical High Energy sources/processes, both extragalactic (like in AGNs, jet-disk symbiosis) or Galactic (like X-ray binaries, XRBs).

In particular in the case of Galactic BH-disk systems, as shown in fig. \ref{fig3}, we see that the temperature can increase to hundreds of keV; we see as well the non-thermal component from very high energy ($> 10^{11}$ eV) observations, but a huge gap (partially bridged by {\it Fermi} and {\it Agile}) in between.

\subsection{Star Forming Regions, Galaxy history and Supernovae}

The experience with {\it Fermi} LAT tells us how important it is to have a good knowledge of the ISM in our Galaxy and especially of the young-active regions ({\it e.g.} \cite{1fgl}).

GRIPS will in fact be a great instrument for studying the history of our Galaxy. There are in fact two unique lines produced after supernovae (SN) explosions that are visible in its energy range: $^{26}$Al (half-life of 700000 years) and $^{44}$Ti (half-life of 60 yrs), both detected by COMPTEL ({\it e.g.} \cite{tial}), that will allow us to map and to study the history of our Galaxy with unprecedented accuracy.

Moreover, regarding SN, another important radioactive line lies in the GRIPS energy range, $^{56}$Ni, that is crucial in order to distinguish SN progenitors and explosion scenarios (within 200 Mpc).

\subsection{Dark Matter}

The recent experience with {\it Fermi} LAT \cite{gc} \cite{gc2} showed how difficult it is to study the Galactic Center (GC) region; in particular, in \cite{gc2} a bump is visible in the energy spectrum at GeV energies that is difficult to evaluate and is compatible with both a population of pulsars and a dark matter (DM) distribution.

Also on this intricate topic, the synergy GeV-MeV can play a crucial role: in fact, the rest mass of the electron, 511 keV, is in the energy range of GRIPS. It allows us to map the positron annihilation from regions such as the GC, or, more in general, from the Galactic bulge.
This will be crucial in order to evaluate the DM component in those regions, that, as seen by LAT (see \cite{morselli} for more details), are not easy to study at all. Moreover having the $\pi_0$bump at 68 MeV will be useful as well in order to be able to map also the hadronic component in those regions (in order to accomplish this task will be useful to increase the energy range upper limit for GRM and this is not difficult to implement in the current configuration of the instrument).

\begin{figure}
\begin{center}
\hbox{
\psfig{figure=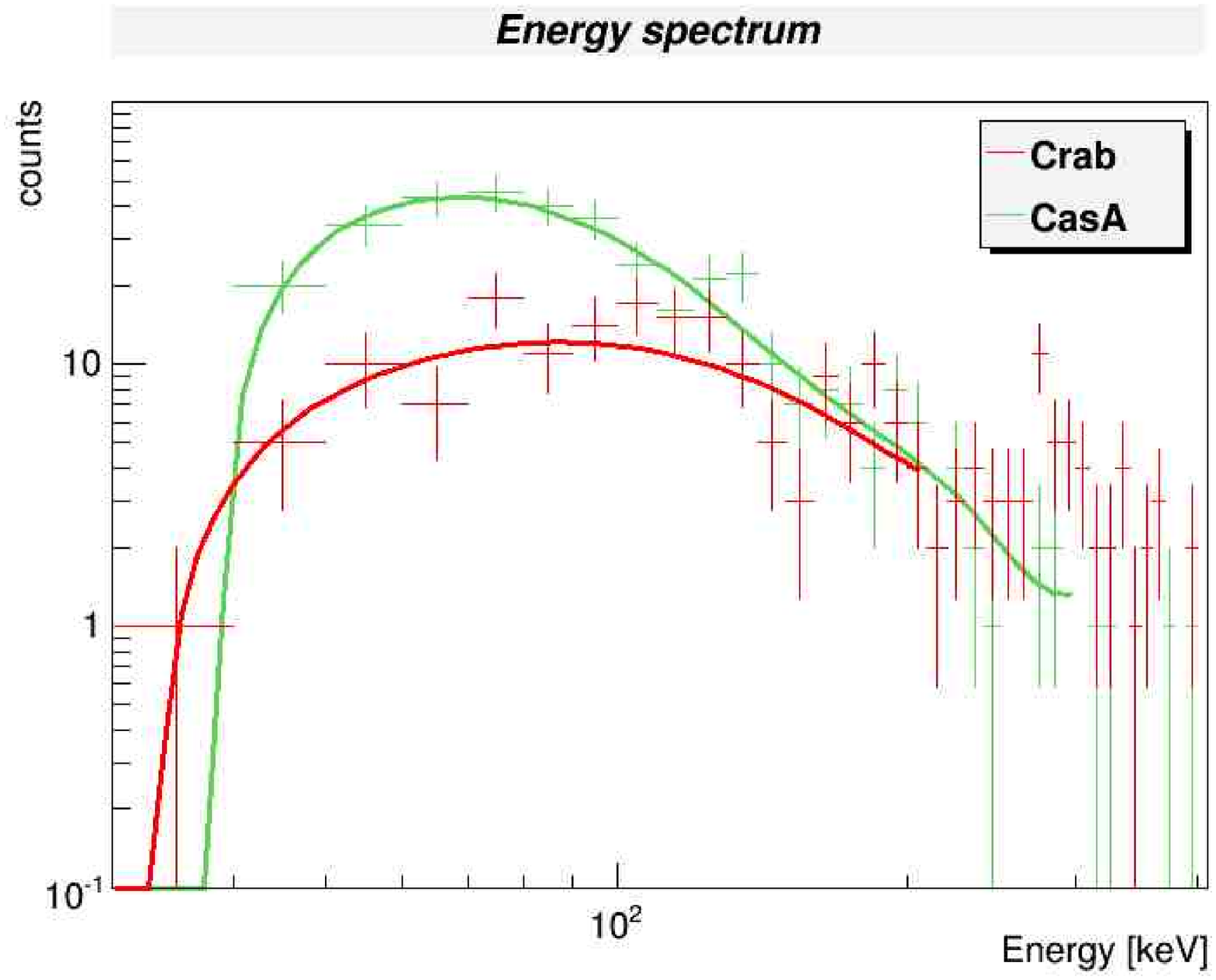,height=5.0cm,angle=0}
\hspace{0.6cm}
\psfig{figure=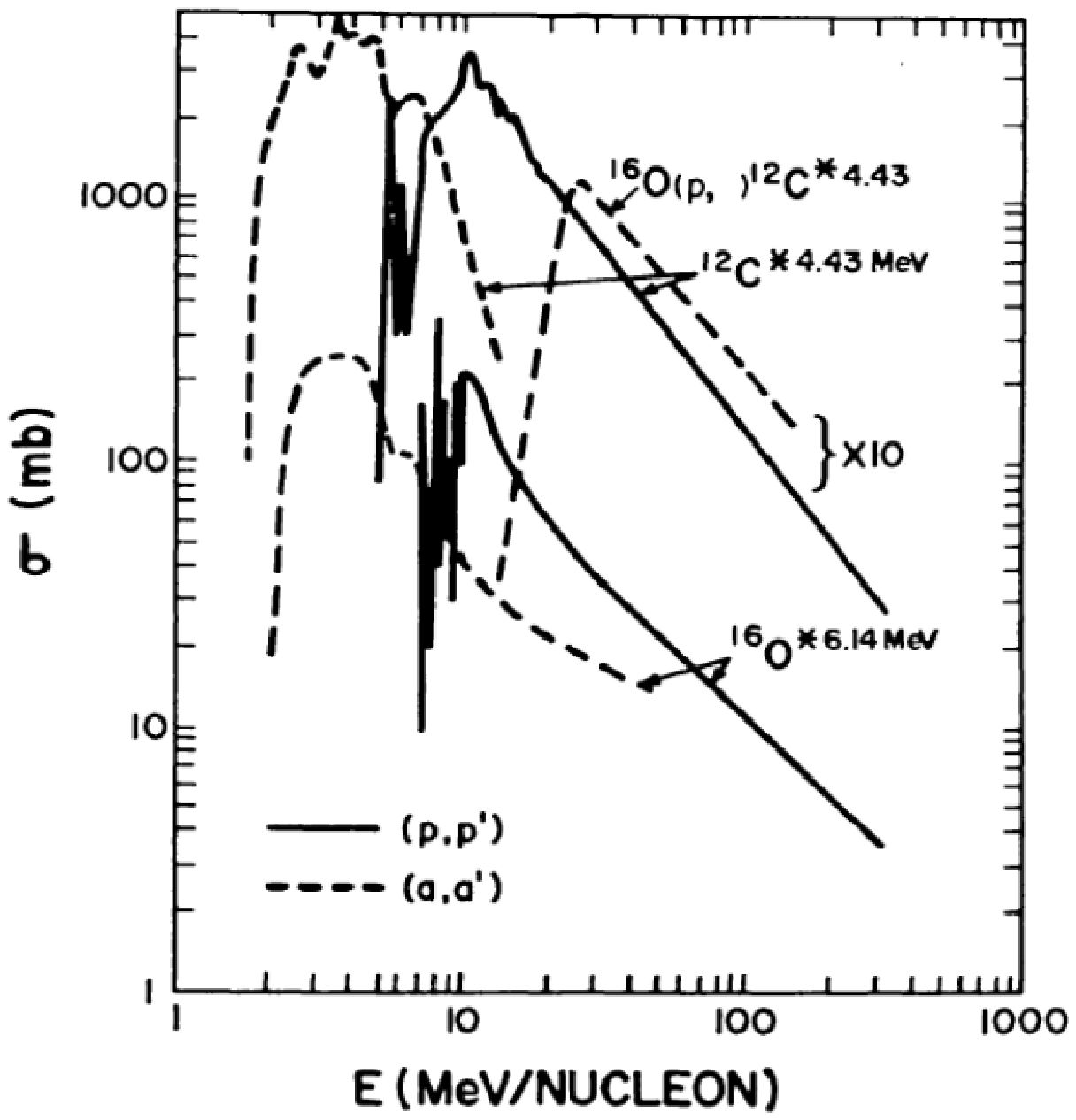,height=5.0cm,angle=0}
\vspace{-1cm}
}
\end{center}
\caption{\footnotesize
{{\it Left panel}: preliminary simulation of 1 minute of {\it GRIPS} observations using MEGAlib software \cite{megalib}: Crab Nebula and Cassiopeia A spectra.
{\it Right panel}: Cross section for the $^{12}$C excitation and (quasi spontaneous) deexcitation line emission at 4.43 MeV, measured in a solar flare of August 1972 \cite{ramaty}.
}
}
\label{fig5}
\end{figure}

\subsection{MeV blazars and extragalactic MeV-GeV gamma-ray background}

There are two types of flat-spectrum radio quasars (FSRQs): those with steep gamma-ray spectrum (MeV blazars) and those with flat gamma-ray spectrum (GeV blazars). 

The two types of blazars are thought to be two aspects of the same phenomenon ({\it e.g.} \cite{sikora}) and it has been predicted that:

- variability timescales of MeV blazars should be longer than variability timescales of GeV blazars;

- both types of blazars are expected to appear in the same object.

Contemporary observations at GeV and MeV energies will be able to answer these questions.

Moreover, \cite{ajello} and \cite{kneise} point out that we can explain only $\sim$16\% of extragalactic MeV-GeV background and consequently need for a numerous faint population of active galactic nuclei (AGNs) $>$ 10 MeV in order to be able to explain the observed isotropic extragalactic gamma-ray diffuse emission.

\subsection{Origin of cosmic rays}

One of the most important achievement of GRIPS will regard cosmic rays (CRs). CRs are $\sim$98\% hadrons and $\sim$2\% leptons, they were discovered  $\sim$1 century ago by Domenico Pacini \cite{crs} and Victor Hess \cite{crs2} and their origin is still unknown.
While the leptons acceleration has been proven (e.g. by X-rays and Radio observations of SNRs or PWNe), the ''smoking gun'' for the hadronic acceleration is still missing. Since the sixties, the SN explosions are thought to be responsible for the hadronic component of CRs \cite{ginzburg} and the recent discoveries at TeV (from Imaging Atmospheric Cherenkov Telescpes, IACTs) and GeV ({\it Fermi} LAT) energies seem to support this, however there is no conclusive proof yet.

GRIPS will provide this proof in two ways:

- by distinguishing between hadronic and leptonic emission from the continuum spectrum.

- by studying the nuclear de-excitation lines.

\subsubsection{Cosmic rays: continuum spectra}

Leptonic and hadronic processes have different signatures in the GRM energy range. {\it GRIPS} will be able to detect the sources that are thought to be responsible of CRs acceleration and to precisely measure their spectra. 

Fig. \ref{fig5} shows the spectra of the Crab Nebula and Cassiopeia A extracted from the preliminary simulation of 1 minute of {\it GRIPS} observations; let us note once more that it would be extremely nice to have for the first time both the 511 keV and the 68 MeV $\pi_0$ bumps in the same instrument.

\subsubsection{Cosmic rays: de-excitation of nuclear lines}

On the other hand, the ''hadronic fingerprint'' can arrive from the study of the de-excitation lines of heavier nuclei: let us take the exemplary case of the carbon line at 4.4 MeV from Cassiopeia A, considering the prediction from the hadronic fit ($p + A \rightarrow \pi_0 + A$) from GeV-TeV data \cite{latcasa}.

Let us take a proton power law $Q_p = Q_0 p^{−2,3}$ with total energy $W_p = \int_{10 \rm{MeV}} Q_p p dp = 4 \times 10^{49} \rm{erg/s}$ from \cite{latcasa} and note that this would correspond to $\sim$2\% of the SNR kinetic energy (less than what is expected in the Ginzburg scenario \cite{voelk} \cite{drury}).

According to \cite{vink}, electrons with a similar slope produce the hard X-rays via synchrotron emission: the Inverse Compton emission is suppressed, owing to a strong magnetic field B $>$ 0.5mG (for leptonic GeV-TeV fits, 0.1 mG is needed).

Considering a heavy ion enriched plasma composition with $n_C = 10 ~\rm{cm}^{−3}$ at reverse shock, the $^{12}$C excitation and (quasi spontaneous) de-excitation line emission at 4.43 MeV is given by:

\begin{figure}
\begin{center}

\psfig{figure=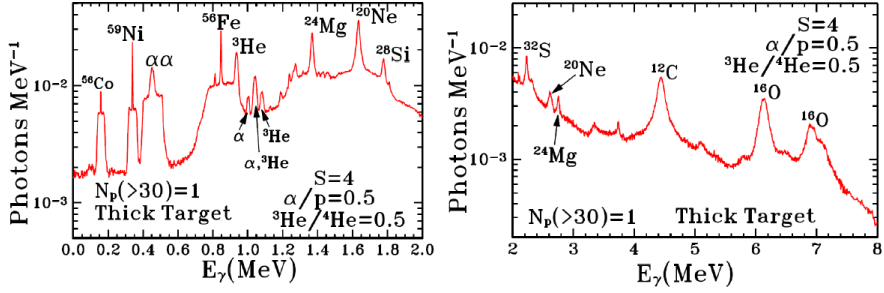,height=4.0cm,angle=0}
\vspace{-0.5cm}

\end{center}
\caption{\footnotesize
{Many other de-excitation lines could be visible in GRM-GRIPS energy range: above all $^{12}$C and $^{16}$O \cite{ramaty}.
}
}
\label{fig6bis}
\end{figure}

\begin{equation}
 \frac{dN}{dE} (^{12}C \rightarrow ^{12}C^{*}) = \int_{10 \rm{MeV}} Q_p (p) \frac{d \sigma (p)}{dp} n_C c dp 
\end{equation}
where the cross section $d \sigma / dp \simeq 3000 \delta (p - 10 \rm{MeV})\rm{mb}$ \cite{ramaty} is shown in fig. \ref{fig5}.

Using $d=3.4$~kpc for the distance to Cassiopeia A, this yields the line flux at 4.43 MeV:

\begin{figure}
\begin{center}

\psfig{figure=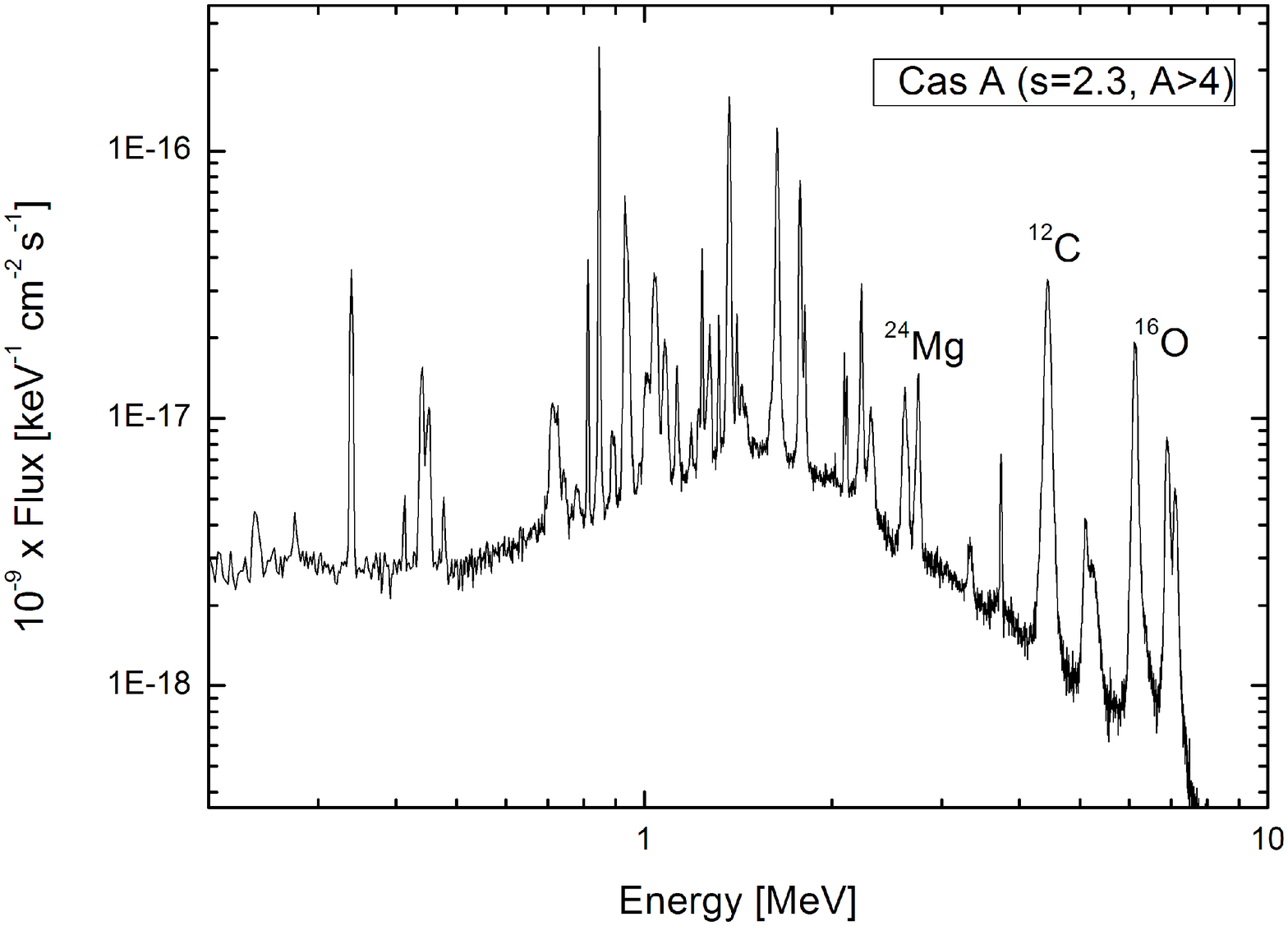,height=5.0cm,angle=0}
\vspace{-1cm}

\end{center}
\caption{\footnotesize
{De-excitation lines for the specific case of Cassiopeia A.
}
}
\label{fig6}
\end{figure}

\begin{equation}
F_{4.43} = \frac{1}{4 \pi d^2} \frac{dN}{dE} \simeq 4 \times 10^{-8} \rm{cm}^{-2} \rm{s}^{-1} \rm{keV}^{-1}
\end{equation}

Considering now a line width of $\Delta E = 100$ keV, we obtain the flux that we were searching for and let us note that such a flux is clearly detectable by GRM-GRIPS; as further check we can also compare it with the upper limit provided by COMPTEL of the flux in the range (3-10) MeV for the continuum \cite{andy}:

\begin{equation}
\Phi_{3-10} = 4 \times 10^{-6} \rm{cm}^{-2} \rm{s}^{-1} < \Phi_{U.L.} = 1.4 \times 10^{-5} \rm{cm}^{-2} \rm{s}^{-1}
\end{equation}

Let's zoom around the $^{12}$C line (fig. \ref{fig6bis}). Using the Ramaty/Koslovsky code \cite{ramaty}, we computed the nuclear de-excitation line spectrum, concluding that with GRIPS we will be able to:

- probe the hadronic acceleration in SNRs in independent ways (e.g.continuum spectrum, $^{12}$C and $^{16}$O lines);

- study the elements abundancies by studying the line ratios.

Using the Ramaty/Koslovsky code, we computed also the nuclear de-excitation line spectrum for the specific case of Cassiopeia A abundancies (visible in fig. \ref{fig6}), noticing the good agreement with the calculation above.

\subsection{Conclusions}

{\it GRIPS} is proposed to be launched in 2020 and its advent will answer many of the open questions in high energy astrophysics; we showed that the answers to such questions are hidden in the MeV gamma-ray band and that deep multi-wavelength studies (in particular together with higher energy instruments, such as {\it Fermi}, {\it Gamma 400} and CTA) will be able to provide the seeked answers.
The deep importance of closing the MeV-gap can be underlined by W. Atwood, one of the fathers of {\it Fermi}: `` If 10 years ago I could do an experiment from hundreds of keV to hundreds of MeV, I would have done it instead of GLAST''. We have shown how crucial {\it GRIPS} will be in answering the one-century old question of the origin of cosmic rays.

\end{document}